\documentclass[3p]{elsarticle}

\usepackage{booktabs}
\usepackage{hepnames}
\usepackage{siunitx}
\usepackage{url}
\usepackage{palatino}
\usepackage[hidelinks,bookmarks=true,plainpages=false]{hyperref}
\usepackage[colorinlistoftodos,prependcaption]{todonotes}

\usepackage{tikz}
\usetikzlibrary{arrows,shapes}
\usetikzlibrary{trees}
\usetikzlibrary{matrix,arrows} 				
\usetikzlibrary{positioning}				
\usetikzlibrary{calc,through}				
\usetikzlibrary{decorations.pathreplacing}  
\usepackage{pgffor}							

\usetikzlibrary{decorations.pathmorphing}	
\usetikzlibrary{decorations.markings}
\tikzset{
	>=stealth', 
    vector/.style={decorate, decoration={snake}, draw},
	provector/.style={decorate, decoration={snake,amplitude=2.5pt}, draw},
	antivector/.style={decorate, decoration={snake,amplitude=-2.5pt}, draw},
    fermion/.style={draw=black, postaction={decorate},
        decoration={markings,mark=at position .55 with {\arrow{>}}}},
    fermionbar/.style={draw=black, postaction={decorate},
        decoration={markings,mark=at position .55 with {\arrow{<}}}},
    fermionnoarrow/.style={draw=black},
    gluon/.style={decorate, draw=black,
        decoration={coil,amplitude=4pt, segment length=5pt}},
    scalar/.style={dashed,draw=black, postaction={decorate},
        decoration={markings,mark=at position .55 with {\arrow{>}}}},
    scalarbar/.style={dashed,draw=black, postaction={decorate},
        decoration={markings,mark=at position .55 with {\arrow{<}}}},
    scalarnoarrow/.style={dashed,draw=black},
    electron/.style={draw=black, postaction={decorate},
        decoration={markings,mark=at position .55 with {\arrow{>}}}},
	bigvector/.style={decorate, decoration={snake,amplitude=4pt}, draw},
}

\definecolor{KIT_gruen_1}{HTML}{00A88F}
\definecolor{KIT_gruen_2}{HTML}{4CC1A5}
\definecolor{KIT_gruen_3}{HTML}{7FD2B8}
\definecolor{KIT_gruen_4}{HTML}{B2E4D1}
\definecolor{KIT_gruen_5}{HTML}{D9F1E6}

\definecolor{KIT_blau_1}{HTML}{4372C1}
\definecolor{KIT_blau_2}{HTML}{7691D2}
\definecolor{KIT_blau_3}{HTML}{9AABDD}
\definecolor{KIT_blau_4}{HTML}{C0C9EA}
\definecolor{KIT_blau_5}{HTML}{E0E3F4}

\definecolor{KIT_maigruen_1}{HTML}{66C42F}
\definecolor{KIT_maigruen_2}{HTML}{93D561}
\definecolor{KIT_maigruen_3}{HTML}{B2E088}
\definecolor{KIT_maigruen_4}{HTML}{D1ECB4}
\definecolor{KIT_maigruen_5}{HTML}{E8F5D8}

\definecolor{KIT_gelb_1}{HTML}{FEE701}
\definecolor{KIT_gelb_2}{HTML}{FEED4B}
\definecolor{KIT_gelb_3}{HTML}{FEF27D}
\definecolor{KIT_gelb_4}{HTML}{FEF7B0}
\definecolor{KIT_gelb_5}{HTML}{FEFBD8}

\definecolor{KIT_orange_1}{HTML}{F69110}
\definecolor{KIT_orange_2}{HTML}{F9AE49}
\definecolor{KIT_orange_3}{HTML}{FAC376}
\definecolor{KIT_orange_4}{HTML}{FCD9A8}
\definecolor{KIT_orange_5}{HTML}{FDECD2}

\definecolor{KIT_braun_1}{HTML}{A97E23}
\definecolor{KIT_braun_2}{HTML}{C09D52}
\definecolor{KIT_braun_3}{HTML}{D0B47A}
\definecolor{KIT_braun_4}{HTML}{E2D0A8}
\definecolor{KIT_braun_5}{HTML}{F0E6D2}

\definecolor{KIT_rot_1}{HTML}{BF2229}
\definecolor{KIT_rot_2}{HTML}{CD574A}
\definecolor{KIT_rot_3}{HTML}{DA806E}
\definecolor{KIT_rot_4}{HTML}{E7AE9D}
\definecolor{KIT_rot_5}{HTML}{F2D5CB}

\definecolor{KIT_lila_1}{HTML}{BC0C8D}
\definecolor{KIT_lila_2}{HTML}{CC4DAE}
\definecolor{KIT_lila_3}{HTML}{D97CC4}
\definecolor{KIT_lila_4}{HTML}{E7AEDB}
\definecolor{KIT_lila_5}{HTML}{F2D6ED}

\definecolor{KIT_cyanblau_1}{HTML}{1CAEEB}
\definecolor{KIT_cyanblau_2}{HTML}{5EC5F1}
\definecolor{KIT_cyanblau_3}{HTML}{8CD4F4}
\definecolor{KIT_cyanblau_4}{HTML}{B9E5F8}
\definecolor{KIT_cyanblau_5}{HTML}{DCF2FB}

\definecolor{KIT_grau_1}{HTML}{231F20}
\definecolor{KIT_grau_2}{HTML}{656263}
\definecolor{KIT_grau_3}{HTML}{918F90}
\definecolor{KIT_grau_4}{HTML}{BDBCBC}
\definecolor{KIT_grau_5}{HTML}{DEDDDE}

\makeatletter
\def\ps@pprintTitle{%
  \let\@oddhead\@empty
  \let\@evenhead\@empty
  \def\@oddfoot{\reset@font\hfil\thepage\hfil}
  \let\@evenfoot\@oddfoot
}
\makeatother

\newcommand{\software}[1]{\textsc{#1}\xspace}
\newcommand{\artus}{\software{Artus}}
\newcommand{\kapa}{{\Large$\kappa$}\software{appa}}
\newcommand{\harryplotter}{\software{HarryPlotter}}

\begin{document}

\begin{frontmatter}
\title{\artus{} -- A Framework for Event-based Data Analysis in High Energy Physics}

\author{Joram Berger}
\author{Fabio Colombo}
\author{Raphael Friese}
\author{Dominik Haitz}
\author{Thomas Hauth}
\author{Thomas M\"uller}
\author{G\"unter Quast}
\author{Georg Sieber}
\address{Karlsruhe Institute of Technology, Wolfgang-Gaede-Stra\ss{}e 1, D-76131 Karlsruhe}

\begin{abstract}
\artus is an event-based data-processing framework for high energy physics experiments.
It is designed for large-scale data analysis in a collaborative environment.
The architecture design choices take into account typical challenges and are based on experiences with similar applications.
The structure of the framework and its advantages are described.
An example use case and performance measurements are presented.
The framework is well-tested and successfully used by several analysis groups.
\end{abstract}

\end{frontmatter}


\section{Typical Workflow of High Energy Physics Analyses}
\label{section_artus_analysis_workflow}
The workflow of the experiments collecting data at the Large Hadron Collider (LHC) at CERN is taken as an example to illustrate the organization and needs of a modern High Energy Physics (HEP) analysis.
Nonetheless, equal or very similar requirements are also present for HEP experiments not located at the LHC.

\begin{figure}[t]
\centering \begin{tikzpicture}[scale=1.0]
\draw[thick, ->] (0.7,0) -- (9.3,0) node[anchor=north] {};
\draw[KIT_gruen_1,->] (0,0) -- (0,4) node[anchor=south] {File size /$\mathrm{MB}$};

\foreach \y/\ytext in {1/0,2/3, 3/6}
\draw[KIT_gruen_1,shift={(0,\y)}] (2pt,0pt)  node[left] {$10^\ytext$};

\filldraw[KIT_gruen_1] (0.3,3.11) circle(.05cm) node[black,anchor=south west]{Full event data};
\filldraw[KIT_gruen_1] (3.3,2.269) circle(.05cm) node[black, anchor=west]{Skim};
\filldraw[KIT_gruen_1] (6.6,1.03) circle(.05cm) node[black, anchor=south west]{N-Tuple};
\filldraw[KIT_gruen_1] (9.7,0) circle(.05cm) node[black, anchor=north]{Plot};

\draw[KIT_rot_1,<-] (10,0) -- (10,4) node[anchor=south] {Run time / h};
\foreach \y/\ytext in {1/-1,2/1, 3/3}
\draw[left,KIT_rot_1, shift={(0,\y)}, anchor=east] (11cm,0pt)  node[left] {\hfill$10^{\ytext}$};

\filldraw[KIT_rot_1] (1.65,0.66) circle(.05cm) node[black, anchor=west]{Skimming};
\filldraw[KIT_rot_1] (5,1.65) circle(.05cm) node[black, anchor=north]{Analysis};
\filldraw[KIT_rot_1] (8.3,3.08) circle(.05cm) node[black, anchor=south]{Plotting};
\draw[KIT_rot_1,->,thick] (0.5,0.3) -- (9.5,3.3) ;
\draw[KIT_gruen_1,->,thick] (.5,3.1) -- (9.5,0.2) ;

\draw (5,0) node[anchor=north] {Analysis Progress};
\end{tikzpicture}
\caption[Characteristics of an event-based analysis in high energy physics]{Characteristic of an event-based analysis in high energy physics, numbers taken from \cite{joram_phd}.
Data size is reduced by six orders of magnitude, the turnaround time by a factor of 10000.
I/O intensive operations are done highly parallelized in the WLCG.
The data reduction is achieved by applying analysis specific event selection steps as early as possible.
Performing the data-intensive operations only in the first step of this sequence reduces the run time of the analysis.}
\label{figure_analysis_steps}
\end{figure}
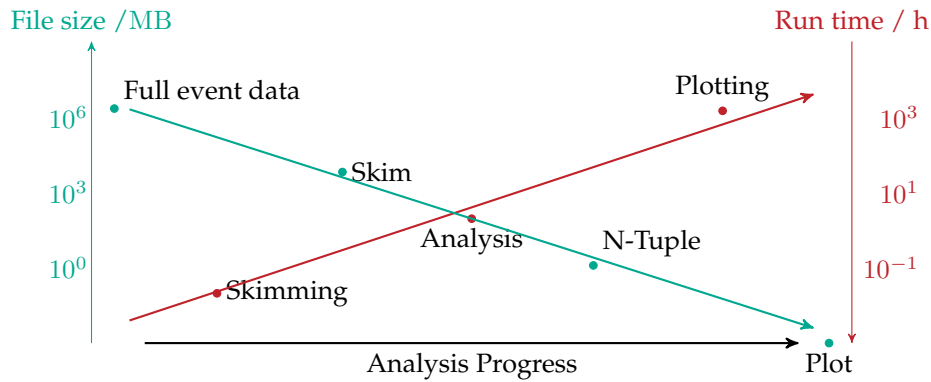

Between the data taking at the detector or the generation of simulated events and the presentation of the physical results, the following analysis steps have shown to be the best compromise in terms of turnaround time and efficiency as explained in Figure \ref{figure_analysis_steps}.

The reconstruction of the raw data coming from the experiments is performed on large computing centers organized in the Worldwide LHC Computing Grid (WLCG)~\cite{wlcg}.
The resulting datasets contain event information and high-level physics objects and are in the order of \SI{100}{TB} in size.

The data which are relevant for a specific analysis are selected in a step usually referred to as \textit{skimming}.
The output of the skimming process is stored in an n-tuple format on local file servers, often owned and operated by a single institute.
The size of these files ranges from several \SI{10}{GB} to few~TB.

Finally, the skimmed n-tuples are processed by user code running on local resources.
This processing step includes further event selections, event categorization or high-level calculations based on the reconstructed physics objects.
The events are usually processed with different configurations, depending on the specific analysis needs.
The output is then used for post-processing steps, including the graphical representation of the results and their statistical interpretation.

In many cases, the code for the last step is developed without a predefined strategy or structure, following the evolving needs of the analysis which are often physics-driven and not always entirely predictable in the code design phase.
This leads to convoluted, tangled and unstructured code, which is difficult to understand and to maintain and also difficult to hand over to new people joining the analysis effort.
Also, code is rarely shared among different groups, although the basic parts of an analysis, such as handling of input data in the same formats, their processing and the selection and identification of physics objects follow similar or equivalent specifications.
Looking at the user side, an analysis group usually consists of several people having different academic levels and different programming skills, some of them joining the analysis effort only for a short period of time and limited to a well-defined task.

\section{Requirements and Challenges for Analysis Software}
\label{section_artus_challenges}
Based on the workflow described in the previous section, the following demands for an analysis framework arise:

\begin{description}
\item[Performance]
The main performance indicator is the \emph{turnaround time} which is the time needed to run the entire analysis in case new requirements arise.
Since the runtime is often limited by the transfer speed of the storage servers, the software should load each event only once and perform all necessary processing steps.
Still, since analysis requirements sometimes change within a few hours, a high level of flexibility is needed.
This explains the need for a modular structure, where only those parts of the code are executed that are really necessary for the analysis.

\item[Maintainability]
Source code maintenance, i.e.\ to be up-to-date with the latest requirements while not losing backwards compatibility, is a challenge for each software project.
To make the entry level as low as possible, beginners can start with an example analysis. 
Up-to-date recipes should allow the installation and execution of the example analysis out-of-the box.
Steering the analysis in detail with configuration files makes it possible to re-use the same code for differing analysis requirements.
The modules should encapsulate the actual operations providing and enforcing a clear code structure while the data is stored in the desired data format.

\item[Reliability]
In most cases, analysis code is written from scratch for very specific tasks and it is used only by very small groups of analysts.
This leads to frequent re-imple\-men\-ta\-tions and low trust in reliability.
An ideal framework should allow to re-use analysis code due to its modular structure as well as sharing it between different analysis groups, and therefore improving code quality significantly.

Source code and configuration files are the most precise form of describing the processes.
Therefore, readability also for people who have not written the code is crucial.
Also, if possible, existing and well-established libraries of the HEP community are interfaced.

\item[Flexibility]
While not aiming at feature-completeness, the software should make it easy to cover all analysis needs by being flexible and extendible.
This comprises new data formats, physical quantites and analysis steps, for instance.
\end{description}

\section{The \artus Framework}
\label{section_artus_motivation}

The framework introduced in this section handles the last step of the workflow, the user analysis.

Following the structure of the CMS software framework~\cite{cms, cmssw}, the \artus framework\footnote{\url{https://github.com/artus-analysis/artus}} provides a front-to-end solution for the analysis of event-based data.
The basic concept of modules processing events within an event loop is complemented by tools for the configuration of every step.
The complete structure is explained in the next section.

This general framework is applicable to any kind of event processing analysis.
The well-tested core is shared among various analyses, providing a reliable fundament for any new analysis.
The highly modular structure guides the user to develop analysis software that is easily maintainable and extensible.
Thus, the \artus framework avoids the aforementioned problems of separate user code for individual analyses.

The \artus framework is written in C++.
The core of the framework only depends on the boost library~\cite{boost}.
Further dependencies may be introduced by the actual analysis code.
C++ is chosen for performance reasons and in order to be able to integrate existing frameworks and libraries that are commonly used in high energy physics.

Multi-core techniques are not required since the processing can be trivially parallelized by splitting the input dataset and sending it to different computing instances.
This is possible because the processed events are known to be independent of each other.
The framework comes with a built-in possibility to submit analysis jobs with the job submission tool grid-control.

\section{Structure of the Framework}
\label{section_artus_structure}

The basic concept of the framework is taken from the code used for the $\PZ+\text{jet}$ analysis documented in~\cite{joram_phd}.
It is generalized and extended to satisfy the needs of various analysis groups.
The structure is illustrated in Figure~\ref{figure_artus_structure}.

\begin{figure}[p]
\centering \begin{tikzpicture}[scale=1.0]

\node[draw, rectangle, rounded corners, minimum width=0.8cm, minimum height=5.3cm, KIT_rot_1] (event) at (0cm-2.5cm-0.5cm,0cm-1cm-2cm+0.75cm) {\rotatebox{90}{Event}};
\node[draw, rectangle, minimum width=0.8cm, minimum height=5.3cm] (eventprovider) at (0cm-2.5cm-0.5cm-1.5cm,0cm-1cm-2cm+0.75cm) {\rotatebox{90}{Event Provider}};
\node[draw, rectangle, rounded corners, minimum width=0.8cm, minimum height=5.3cm, KIT_rot_1, fill=KIT_rot_5] (input) at (0cm-2.5cm-0.5cm-1.5cm-1.5cm,0cm-1cm-2cm+0.75cm) {\rotatebox{90}{Input}};

\node[draw, rectangle, minimum width=3.8cm, minimum height=0.8cm] (globalpipeline) at (0cm,0cm) {Global Pipeline};

\draw[->, thick] ($ (input) + (0.4cm,2.2cm) $) -- ($ (eventprovider) + (-0.4cm,2.2cm) $);
\draw[->, thick] ($ (eventprovider) + (0.4cm,2.2cm) $) -- ($ (event) + (-0.4cm,2.2cm) $);
\draw[->, thick] ($ (event) + (0.4cm,2.2cm) $) -- (globalpipeline);

\draw[->, thick, KIT_rot_1] ($ (event) + (0.4cm,2.4cm) $) -- ($ (event) + (0.4cm+0.9cm,2.4cm) $);
\draw[->, thick, KIT_rot_1] ($ (event) + (0.4cm,0.5cm) $) -- ($ (event) + (0.4cm+0.9cm,0.5cm) $);
\draw[->, thick, KIT_rot_1] ($ (event) + (0.4cm,0.4cm) $) -- ($ (event) + (0.4cm+1.9cm,0.4cm) $);
\draw[->, thick, KIT_rot_1] ($ (event) + (0.4cm,0.3cm) $) -- ($ (event) + (0.4cm+2.9cm,0.3cm) $);

\node[draw, rectangle, minimum width=0.8cm, minimum height=3.3cm] (localpipeline1) at (0cm-1.5cm,0cm-1cm-2cm-0.25cm) {\rotatebox{90}{Local Pipeline 1}};
\node[draw, rectangle, minimum width=0.8cm, minimum height=3.3cm] (localpipeline2) at (0cm-0.5cm,0cm-1cm-2cm-0.25cm) {\rotatebox{90}{Local Pipeline 2}};
\node[draw, rectangle, minimum width=0.8cm, minimum height=3.3cm] (localpipeline3) at (0cm+0.5cm,0cm-1cm-2cm-0.25cm) {\rotatebox{90}{Local Pipeline 3}};
\node[rectangle, minimum width=0.8cm, minimum height=3.3cm] (localpipeline4) at (0cm+1.5cm,0cm-1cm-2cm-0.25cm) {\textbf{$\cdots$}};
\node[KIT_rot_1] at ($ (globalpipeline) + (-0.5cm+0.5cm,0cm-1cm-0.2cm+1.5ex) $) {\textit{duplicate product}};

\draw[->, thick] (globalpipeline) -- ($ (globalpipeline) + (2cm+0.2cm,0cm) $) -- ($ (globalpipeline) + (2cm+0.2cm,0cm-1cm-0.2cm) $) -- ($ (globalpipeline) + (-2cm+0.5cm,0cm-1cm-0.2cm) $) -- (localpipeline1);
\draw[->, thick] (globalpipeline) -- ($ (globalpipeline) + (2cm+0.2cm,0cm) $) -- ($ (globalpipeline) + (2cm+0.2cm,0cm-1cm-0.2cm) $) -- ($ (globalpipeline) + (-2cm+1.5cm,0cm-1cm-0.2cm) $) -- (localpipeline2);
\draw[->, thick] (globalpipeline) -- ($ (globalpipeline) + (2cm+0.2cm,0cm) $) -- ($ (globalpipeline) + (2cm+0.2cm,0cm-1cm-0.2cm) $) -- ($ (globalpipeline) + (-2cm+2.5cm,0cm-1cm-0.2cm) $) -- (localpipeline3);

\node[draw, rectangle, rounded corners, minimum width=3.8cm, minimum height=0.8cm, KIT_blau_1, fill=KIT_blau_5] (output) at (0cm,0cm-1cm-2cm-2cm-1cm) {Output};
\draw[dotted, thick, KIT_blau_1] ($ (output) + (-1cm,0.4cm) $) -- ($ (output) + (-1cm,-0.4cm) $);
\draw[dotted, thick, KIT_blau_1] ($ (output) + (0cm,0.4cm) $) -- ($ (output) + (0cm,-0.4cm) $);
\draw[dotted, thick, KIT_blau_1] ($ (output) + (1cm,0.4cm) $) -- ($ (output) + (1cm,-0.4cm) $);

\draw[->, thick] ($ (localpipeline1) + (0cm,-1.65cm) $) -- ($ (localpipeline1) + (0cm,-1.75cm-0.5cm-0.1cm) $);
\draw[->, thick] ($ (localpipeline2) + (0cm,-1.65cm) $) -- ($ (localpipeline2) + (0cm,-1.75cm-0.5cm-0.1cm) $);
\draw[->, thick] ($ (localpipeline3) + (0cm,-1.65cm) $) -- ($ (localpipeline3) + (0cm,-1.75cm-0.5cm-0.1cm) $);

\node[draw, rectangle, rounded corners, minimum width=0.8cm, minimum height=5.3cm, KIT_gruen_1, fill=KIT_gruen_5] (configuration) at (0cm+2.5cm+0.5cm,0cm-1cm-2cm+0.75cm) {\rotatebox{90}{Configuration}};

\draw[dotted, thick, KIT_gruen_1] ($ (configuration) + (-0.4cm,1.5cm) $) -- ($ (configuration) + (0.4cm,1.5cm) $);

\draw[->, thick, KIT_gruen_1] ($ (configuration) + (-0.4cm,2.4cm) $) -- ($ (configuration) + (-0.4cm-0.9cm,2.4cm) $);
\draw[->, thick, KIT_gruen_1] ($ (configuration) + (-0.4cm,0.5cm) $) -- ($ (configuration) + (-0.4cm-3.9cm,0.5cm) $);
\draw[->, thick, KIT_gruen_1] ($ (configuration) + (-0.4cm,0.4cm) $) -- ($ (configuration) + (-0.4cm-2.9cm,0.4cm) $);
\draw[->, thick, KIT_gruen_1] ($ (configuration) + (-0.4cm,0.3cm) $) -- ($ (configuration) + (-0.4cm-1.9cm,0.3cm) $);
\draw[->, thick, KIT_gruen_1] (configuration) -- ($ (configuration) + (0cm,2.65cm+0.4cm) $) -- ($ (eventprovider) + (-0.2cm,2.65cm+0.4cm) $) -- ($ (eventprovider) + (-0.2cm,2.65cm-0.2cm) $);

\draw[->, thick] (configuration) |- (output);

\draw[dashed, thin] ($ (eventprovider) + (0cm,2.65cm) $) -- ($ (eventprovider) + (0cm,2.65cm+0.2cm) $) -- ($ (configuration) + (-0.6cm,2.65cm+0.2cm) $) -- ($ (configuration) + (-0.6cm,-2.65cm-0.2cm) $) -- ($ (eventprovider) + (0cm,-2.65cm-0.2cm) $) -- ($ (eventprovider) + (0cm,-2.65cm) $);
\node at ($ (eventprovider) + (1cm,-2.65cm-0.2cm-1.5ex) $) {\textit{Event Loop}};
\end{tikzpicture}
\caption[Structure of an \artus analysis.]{Structure of an \artus analysis.
The input is read by an event provider.
Within the pipelines the event content is analyzed by the processors.
Consumers in local pipelines write results to a common output.
All parts of the analysis are configurable.}
\label{figure_artus_structure}

\vspace{10ex}

\centering \begin{tikzpicture}[scale=1.0]
\node at (0cm,0cm) [draw, rectangle, minimum height=2.5cm, minimum width=5ex, node distance=2.5cm+0.2cm] (processor1) {\rotatebox{90}{Producer A}};
\node [draw, rectangle, minimum height=2.5cm, minimum width=5ex, node distance=5ex+0.3cm, right of=processor1] (processor2) {\rotatebox{90}{Filter B}};
\node [draw, rectangle, minimum height=2.5cm, minimum width=5ex, node distance=5ex+0.3cm, right of=processor2] (processor3) {\rotatebox{90}{Producer C}};
\node [draw, rectangle, minimum height=2.5cm, minimum width=5ex, node distance=5ex+0.3cm, right of=processor3] (processor4) {\rotatebox{90}{Filter D}};
\node [draw, rectangle, minimum height=2.5cm, minimum width=5ex, node distance=5ex+0.3cm, right of=processor4] (processor5) {\rotatebox{90}{Filter E}};
\node [draw, rectangle, minimum height=2.5cm, minimum width=5ex, node distance=5ex+0.3cm, right of=processor5] (processor6) {\rotatebox{90}{Producer F}};
\node [draw, rectangle, minimum height=2.5cm, minimum width=5ex, node distance=5ex+0.3cm, right of=processor6] (processor7) {\rotatebox{90}{Producer G}};
\node [draw, rectangle, minimum height=2.5cm, minimum width=5ex, node distance=5ex+0.3cm, right of=processor7] (processor8) {\rotatebox{90}{Producer H}};
\node [rectangle, minimum height=2.5cm, minimum width=5ex, node distance=5ex+0.3cm, right of=processor8] (processor9) {\textbf{$\cdots$}};
\node [draw, rectangle, minimum height=2.5cm, minimum width=5ex, node distance=5ex+0.3cm, right of=processor9] (processor10) {\rotatebox{90}{Consumer 1}};
\node [draw, rectangle, minimum height=2.5cm, minimum width=5ex, node distance=5ex+0.3cm, right of=processor10] (processor11) {\rotatebox{90}{Consumer 2}};
\node [draw, rectangle, minimum height=2.5cm, minimum width=5ex, node distance=5ex+0.3cm, right of=processor11] (processor12) {\rotatebox{90}{Consumer 3}};
\node [rectangle, minimum height=2.5cm, minimum width=5ex, node distance=5ex+0.3cm, right of=processor12] (processor13) {\textbf{$\cdots$}};

\draw[->, thick] (processor1) -- (processor2);
\draw[->, thick] (processor2) -- (processor3);
\draw[->, thick] (processor3) -- (processor4);
\draw[->, thick] (processor4) -- (processor5);
\draw[->, thick] (processor5) -- (processor6);
\draw[->, thick] (processor6) -- (processor7);
\draw[->, thick] (processor7) -- (processor8);
\draw[->, thick] (processor8) -- (processor9);
\draw[->, thick] (processor9) -- (processor10);
\draw[->, thick] (processor10) -- (processor11);
\draw[->, thick] (processor11) -- (processor12);
\draw[->, thick] (processor12) -- (processor13);

\node [draw, rectangle, rounded corners, minimum width=3.5cm, minimum height=5ex, node distance=1.25cm+2.5ex+1cm, above of=processor5, KIT_gruen_1] (settings) {Settings};
\node [draw, rectangle, rounded corners, minimum width=3.5cm, minimum height=5ex, node distance=1.25cm+2.5ex+1cm+5ex+0.2cm, above of=processor9, KIT_rot_1] (event) {Event};
\node [draw, rectangle, rounded corners, minimum width=3.5cm, minimum height=5ex, node distance=1.25cm+2.5ex+1cm, below of=processor5, KIT_rot_1] (product) {Product};
\node [draw, rectangle, rounded corners, minimum width=3.5cm, minimum height=5ex, node distance=1.25cm+2.5ex+1cm+5ex-0.05cm, below of=processor11, KIT_blau_1] (output) {Output};

\draw[->, thick, KIT_gruen_1] (settings) -- ($ (settings) + (0cm,-2.5ex-0.4cm) $) -- ($ (processor1) + (-1ex,1.25cm+0.6cm) $) -- ($ (processor1) + (-1ex,1.25cm) $);
\draw[->, thick, KIT_gruen_1] (settings) -- ($ (settings) + (0cm,-2.5ex-0.4cm) $) -- ($ (processor2) + (-1ex,1.25cm+0.6cm) $) -- ($ (processor2) + (-1ex,1.25cm) $);
\draw[->, thick, KIT_gruen_1] (settings) -- ($ (settings) + (0cm,-2.5ex-0.4cm) $) -- ($ (processor3) + (-1ex,1.25cm+0.6cm) $) -- ($ (processor3) + (-1ex,1.25cm) $);
\draw[->, thick, KIT_gruen_1] (settings) -- ($ (settings) + (0cm,-2.5ex-0.4cm) $) -- ($ (processor4) + (-1ex,1.25cm+0.6cm) $) -- ($ (processor4) + (-1ex,1.25cm) $);
\draw[->, thick, KIT_gruen_1] (settings) -- ($ (settings) + (0cm,-2.5ex-0.4cm) $) -- ($ (processor5) + (-1ex,1.25cm+0.6cm) $) -- ($ (processor5) + (-1ex,1.25cm) $);
\draw[->, thick, KIT_gruen_1] (settings) -- ($ (settings) + (0cm,-2.5ex-0.4cm) $) -- ($ (processor6) + (-1ex,1.25cm+0.6cm) $) -- ($ (processor6) + (-1ex,1.25cm) $);
\draw[->, thick, KIT_gruen_1] (settings) -- ($ (settings) + (0cm,-2.5ex-0.4cm) $) -- ($ (processor7) + (-1ex,1.25cm+0.6cm) $) -- ($ (processor7) + (-1ex,1.25cm) $);
\draw[->, thick, KIT_gruen_1] (settings) -- ($ (settings) + (0cm,-2.5ex-0.4cm) $) -- ($ (processor8) + (-1ex,1.25cm+0.6cm) $) -- ($ (processor8) + (-1ex,1.25cm) $);
\draw[->, thick, KIT_gruen_1] (settings) -- ($ (settings) + (0cm,-2.5ex-0.4cm) $) -- ($ (processor10) + (-1ex,1.25cm+0.6cm) $) -- ($ (processor10) + (-1ex,1.25cm) $);
\draw[->, thick, KIT_gruen_1] (settings) -- ($ (settings) + (0cm,-2.5ex-0.4cm) $) -- ($ (processor11) + (-1ex,1.25cm+0.6cm) $) -- ($ (processor11) + (-1ex,1.25cm) $);
\draw[->, thick, KIT_gruen_1] (settings) -- ($ (settings) + (0cm,-2.5ex-0.4cm) $) -- ($ (processor12) + (-1ex,1.25cm+0.6cm) $) -- ($ (processor12) + (-1ex,1.25cm) $);

\draw[->, thick, KIT_rot_1] (event) -- ($ (event) + (0cm,-5ex-0.2cm-2.5ex-0.6cm) $) -- ($ (processor1) + (1ex,1.25cm+0.4cm) $) -- ($ (processor1) + (1ex,1.25cm) $);
\draw[->, thick, KIT_rot_1] (event) -- ($ (event) + (0cm,-5ex-0.2cm-2.5ex-0.6cm) $) -- ($ (processor2) + (1ex,1.25cm+0.4cm) $) -- ($ (processor2) + (1ex,1.25cm) $);
\draw[->, thick, KIT_rot_1] (event) -- ($ (event) + (0cm,-5ex-0.2cm-2.5ex-0.6cm) $) -- ($ (processor3) + (1ex,1.25cm+0.4cm) $) -- ($ (processor3) + (1ex,1.25cm) $);
\draw[->, thick, KIT_rot_1] (event) -- ($ (event) + (0cm,-5ex-0.2cm-2.5ex-0.6cm) $) -- ($ (processor4) + (1ex,1.25cm+0.4cm) $) -- ($ (processor4) + (1ex,1.25cm) $);
\draw[->, thick, KIT_rot_1] (event) -- ($ (event) + (0cm,-5ex-0.2cm-2.5ex-0.6cm) $) -- ($ (processor5) + (1ex,1.25cm+0.4cm) $) -- ($ (processor5) + (1ex,1.25cm) $);
\draw[->, thick, KIT_rot_1] (event) -- ($ (event) + (0cm,-5ex-0.2cm-2.5ex-0.6cm) $) -- ($ (processor6) + (1ex,1.25cm+0.4cm) $) -- ($ (processor6) + (1ex,1.25cm) $);
\draw[->, thick, KIT_rot_1] (event) -- ($ (event) + (0cm,-5ex-0.2cm-2.5ex-0.6cm) $) -- ($ (processor7) + (1ex,1.25cm+0.4cm) $) -- ($ (processor7) + (1ex,1.25cm) $);
\draw[->, thick, KIT_rot_1] (event) -- ($ (event) + (0cm,-5ex-0.2cm-2.5ex-0.6cm) $) -- ($ (processor8) + (1ex,1.25cm+0.4cm) $) -- ($ (processor8) + (1ex,1.25cm) $);
\draw[->, thick, KIT_rot_1] (event) -- ($ (event) + (0cm,-5ex-0.2cm-2.5ex-0.6cm) $) -- ($ (processor10) + (1ex,1.25cm+0.4cm) $) -- ($ (processor10) + (1ex,1.25cm) $);
\draw[->, thick, KIT_rot_1] (event) -- ($ (event) + (0cm,-5ex-0.2cm-2.5ex-0.6cm) $) -- ($ (processor11) + (1ex,1.25cm+0.4cm) $) -- ($ (processor11) + (1ex,1.25cm) $);
\draw[->, thick, KIT_rot_1] (event) -- ($ (event) + (0cm,-5ex-0.2cm-2.5ex-0.6cm) $) -- ($ (processor12) + (1ex,1.25cm+0.4cm) $) -- ($ (processor12) + (1ex,1.25cm) $);

\draw[<->, very thick, dashed, KIT_rot_1] ($ (product) + (-1.3cm,2.5ex) $) -- ($ (product) + (-1.3cm,2.5ex+0.4cm) $) -- ($ (processor1) + (0cm,-1.25cm-0.6cm) $) -- (processor1);
\draw[<->, very thick, dashed, KIT_rot_1] ($ (product) + (-1.3cm,2.5ex) $) -- ($ (product) + (-1.3cm,2.5ex+0.4cm) $) -- ($ (processor3) + (0cm,-1.25cm-0.6cm) $) -- (processor3);
\draw[<->, very thick, dashed, KIT_rot_1] ($ (product) + (-1.3cm,2.5ex) $) -- ($ (product) + (-1.3cm,2.5ex+0.4cm) $) -- ($ (processor6) + (0cm,-1.25cm-0.6cm) $) -- (processor6);
\draw[<->, very thick, dashed, KIT_rot_1] ($ (product) + (-1.3cm,2.5ex) $) -- ($ (product) + (-1.3cm,2.5ex+0.4cm) $) -- ($ (processor7) + (0cm,-1.25cm-0.6cm) $) -- (processor7);
\draw[<->, very thick, dashed, KIT_rot_1] ($ (product) + (-1.3cm,2.5ex) $) -- ($ (product) + (-1.3cm,2.5ex+0.4cm) $) -- ($ (processor8) + (0cm,-1.25cm-0.6cm) $) -- (processor8);

\draw[->, thick, KIT_rot_1] ($ (product) + (1.3cm,2.5ex) $) -- ($ (product) + (1.3cm,2.5ex+0.6cm) $) -- ($ (processor2) + (0cm,-1.25cm-0.4cm) $) -- (processor2);
\draw[->, thick, KIT_rot_1] ($ (product) + (1.3cm,2.5ex) $) -- ($ (product) + (1.3cm,2.5ex+0.6cm) $) -- ($ (processor4) + (0cm,-1.25cm-0.4cm) $) -- (processor4);
\draw[->, thick, KIT_rot_1] ($ (product) + (1.3cm,2.5ex) $) -- ($ (product) + (1.3cm,2.5ex+0.6cm) $) -- ($ (processor5) + (0cm,-1.25cm-0.4cm) $) -- (processor5);
\draw[->, thick, KIT_rot_1] ($ (product) + (1.3cm,2.5ex) $) -- ($ (product) + (1.3cm,2.5ex+0.6cm) $) -- ($ (processor10) + (-1ex,-1.25cm-0.4cm) $) -- ($ (processor10) + (-1ex,-1.25cm) $);
\draw[->, thick, KIT_rot_1] ($ (product) + (1.3cm,2.5ex) $) -- ($ (product) + (1.3cm,2.5ex+0.6cm) $) -- ($ (processor11) + (-1ex,-1.25cm-0.4cm) $) -- ($ (processor11) + (-1ex,-1.25cm) $);
\draw[->, thick, KIT_rot_1] ($ (product) + (1.3cm,2.5ex) $) -- ($ (product) + (1.3cm,2.5ex+0.6cm) $) -- ($ (processor12) + (-1ex,-1.25cm-0.4cm) $) -- ($ (processor12) + (-1ex,-1.25cm) $);

\draw[<-, thick, KIT_blau_1] ($ (output) + (1ex,2.5ex) $) -- ($ (output) + (1ex,2.5ex+0.4cm+5ex-0.05cm) $) -- ($ (processor10) + (1ex,-1.25cm-0.6cm) $) -- ($ (processor10) + (1ex,-1.25cm) $);
\draw[<-, thick, KIT_blau_1] ($ (output) + (1ex,2.5ex) $) -- ($ (output) + (1ex,2.5ex+0.4cm+5ex-0.05cm) $) -- ($ (processor11) + (1ex,-1.25cm-0.6cm) $) -- ($ (processor11) + (1ex,-1.25cm) $);
\draw[<-, thick, KIT_blau_1] ($ (output) + (1ex,2.5ex) $) -- ($ (output) + (1ex,2.5ex+0.4cm+5ex-0.05cm) $) -- ($ (processor12) + (1ex,-1.25cm-0.6cm) $) -- ($ (processor12) + (1ex,-1.25cm) $);

\draw[dashed, thin] ($ (processor1) + (-2.5ex-0.2cm,1.25cm+1cm+5ex+0.1cm) $) rectangle ($ (processor13) + (2.5ex+0.2cm,-1.25cm-1cm-5ex-0.1cm) $);
\node at ($ (processor1) + (-2.5ex-0.2cm+2em,1.25cm+1cm+5ex+0.1cm+1.5ex) $) {\textit{Pipeline}};
\end{tikzpicture}
\caption[Organization of processors in a pipeline.]{Organization of processors in a pipeline.
Producers and filters are run before the consumers.
All processors can access the pipeline settings, the event content and the product.
Only producers have write access to the product.
Consumers can write results to the output file.}
\label{figure_artus_pipeline}
\end{figure}

\artus respects the separation of code and configuration.
Settings specify what has to be done and with which parameters.
The code performs the selected steps---ideally without requiring the user to be fully aware of implementation details.
Hence, the JSON file completely defines the analysis.
It contains information on the input data and settings for the pipelines and processors described below.
Each of the modules has access to the settings read in from the configuration file.
The \artus package comes with sophisticated tools written in Python which help in generating this JSON file.

The input is handled by an event provider.
Its purpose is to manage the import of event contents from any kind of input to the internal event object and to provide information about the total number of events to be processed.

The events are then sequentially processed by processors organized in pipelines.
First, a so-called global pipeline is run.
Then the execution is split up into multiple so-called local pipelines which are also processed sequentially.
The composition of a pipeline is illustrated in Figure~\ref{figure_artus_pipeline}.

Each pipeline has its own settings.
The local pipelines can access both their own settings and the ones of the global pipeline.
Thus, a mechanism exists that enables processing the same event with different configurations, whereas the costly reading of the input is only done once.
Similarly, a product object is managed by each pipeline.
The product is meant to store new quantities produced in the analysis.
The product for the global pipeline is empty at the beginning of the processing of a new event.
Local pipelines start with a copy of the global product.

A pipeline consists of an arbitrary number of processor modules that are run in a sequence.
Three kinds of processors are distinguished:
\begin{description}
\item[Producers] are meant to calculate new quantities based on the information in the event and the product according to the settings of the pipeline.
\item[Filters] apply a selection based on event properties.
In case an event does not fulfill a requirement defined by a filter, the execution of the subsequent processors can be skipped.
\item[Consumers] can be added to local pipelines.
They are executed after the producers and filters.
The task of consumers is to take quantities from the event and the product and store them in the output according to the pipeline settings.
\end{description}
Producers and filters can be configured in an arbitrary order.
Consumers are executed after the chain of producers and filters.
The output together with the complete configuration is written to a ROOT~\cite{root} file.
These processors are created and managed by a dedicated factory class according to the configuration.

This structure extends the established concepts of existing frameworks~\cite{cmssw} to the following four main advantages:
\begin{description}
\item[Modular structure] Analysis steps are implemented in processors, breaking down the complete analysis in well-defined and easily understandable and writable parts.
\item[Re-utilization of code] Processors can be shared among different analysis groups.
Every user contributes to the testing of existing code and improves the confidence in it.
\item[Configurability] Each module uses settings which are read from the configuration.
A separate sub-set of the configuration is created for each pipeline.
\item[High performance] The concept of pipelines allows to analyze the data with different configurations at the same time, while the data is read in only once and common processing steps are shared in the global pipeline.
Therefore, the cost-intensive input operations are reduced to a minimum.
\end{description}

\section{Building an End-user Analysis}
\label{section_artus_analysis}
In order to create a custom analysis, the user can build his own code on the \artus classes by deriving from them.
He must provide an implementation of the settings, the event and the product class.
The settings class defines the tags that are read from the configuration file and the types of the corresponding values.
The \emph{event} class defines the event content which is read from the input and the \emph{product} class contains members for new quantities calculated during the analysis stage.
These classes need to be derived from the base classes provided by the core of the \artus framework.

An implementation of an \emph{event provider} must be available.
The event provider must know how to read event content from the input files and how to store them in the event object.
Furthermore, a factory class defining the creation of analysis-specific processors is required.
These classes also have to be derived from the \artus base classes.

The most important part of an analysis are the \emph{processors}.
The abstract processors of the \artus core need to be implemented to realize the physics needs of the analysis.
After the basic structure of an analysis is set up, most changes and additions are only related to the processors.

Analyses shareing the input file format can easily also share processors which implement basic analysis steps in a general fashion.
Analysis-specific parts can either derive from existing processor classes or be modeled in new processors.

\section{\harryplotter -- A Python Framework for Post-processing}
\label{section_artus_harryplotter}

The computationally expensive event processing steps are followed by the post-processing of the results and their presentation.
From a technical point of view, the focus here is rather on flexibility than on performance.
The inputs for this step are either histograms filled in the previous step or n-tuples whose entries can be filled into histograms with reasonable computing costs.
For this reason, Python has been chosen as language for the \harryplotter framework\footnote{\url{https://github.com/artus-analysis/Artus/tree/master/HarryPlotter}}.
As the name implies, the main focus is on plotting of results.
The modular structure allows to integrate multiple analysis steps and several output formats.

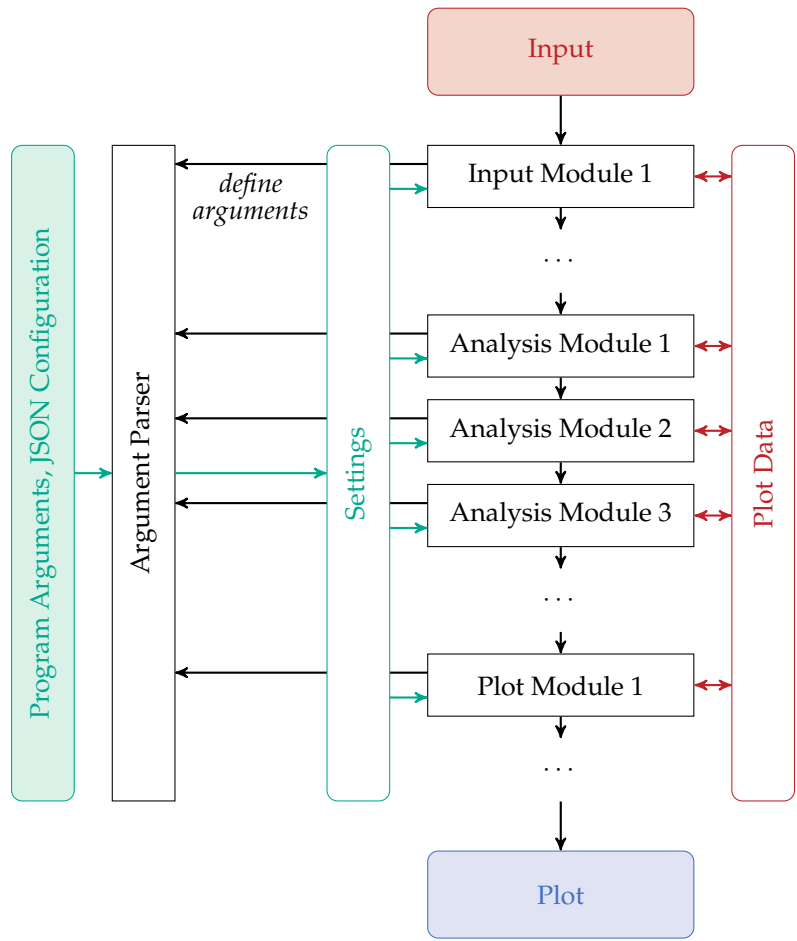
\begin{figure}[htb]
\centering \begin{tikzpicture}[scale=1.0]
\node at (0cm,0cm) [draw, rectangle, minimum width=3.5cm, minimum height=5ex, node distance=5ex+0.3cm] (module1) {Input Module 1};
\node [rectangle, fill=white, minimum width=3.5cm, minimum height=5ex, node distance=5ex+0.3cm, below of=module1] (module2) {$\cdots$};
\node [draw, rectangle, fill=white, minimum width=3.5cm, minimum height=5ex, node distance=5ex+0.3cm, below of=module2] (module3) {Analysis Module 1};
\node [draw, rectangle, fill=white, minimum width=3.5cm, minimum height=5ex, node distance=5ex+0.3cm, below of=module3] (module4) {Analysis Module 2};
\node [draw, rectangle, fill=white, minimum width=3.5cm, minimum height=5ex, node distance=5ex+0.3cm, below of=module4] (module5) {Analysis Module 3};
\node [rectangle, fill=white, minimum width=3.5cm, minimum height=5ex, node distance=5ex+0.3cm, below of=module5] (module6) {$\cdots$};
\node [draw, rectangle, fill=white, minimum width=3.5cm, minimum height=5ex, node distance=5ex+0.3cm, below of=module6] (module7) {Plot Module 1};
\node [rectangle, fill=white, minimum width=3.5cm, minimum height=5ex, node distance=5ex+0.3cm, below of=module7] (module8) {$\cdots$};

\node [draw, rectangle, rounded corners, KIT_rot_1, fill=KIT_rot_5, minimum width=3.5cm, minimum height=7ex, node distance=7ex+0.5cm, above of=module1] (input) {Input};
\node [draw, rectangle, rounded corners, KIT_blau_1, fill=KIT_blau_5, minimum width=3.5cm, minimum height=7ex, node distance=7ex+0.5cm, below of=module8] (plot) {Plot};

\draw [->, thick] (input) -- (module1);
\draw [->, thick] (module1) -- (module2);
\draw [->, thick] (module2) -- (module3);
\draw [->, thick] (module3) -- (module4);
\draw [->, thick] (module4) -- (module5);
\draw [->, thick] (module5) -- (module6);
\draw [->, thick] (module6) -- (module7);
\draw [->, thick] (module7) -- (module8);
\draw [->, thick] (module8) -- (plot);

\draw [fill=white] ($ (module1) + (-1.75cm-0.5cm-5ex-0.5cm-5ex-1.5cm,2.5ex) $) rectangle node{\rotatebox{90}{Argument Parser}} ($ (module8) + (-1.75cm-0.5cm-5ex-0.5cm-1.5cm,-2.5ex) $);

\draw [->, thick] ($ (module1) + (-1.75cm,1ex) $) -- ($ (module1) + (-1.75cm-0.5cm-5ex-0.5cm-1.5cm,1ex) $);
\draw [->, thick] ($ (module3) + (-1.75cm,1ex) $) -- ($ (module3) + (-1.75cm-0.5cm-5ex-0.5cm-1.5cm,1ex) $);
\draw [->, thick] ($ (module4) + (-1.75cm,1ex) $) -- ($ (module4) + (-1.75cm-0.5cm-5ex-0.5cm-1.5cm,1ex) $);
\draw [->, thick] ($ (module5) + (-1.75cm,1ex) $) -- ($ (module5) + (-1.75cm-0.5cm-5ex-0.5cm-1.5cm,1ex) $);
\draw [->, thick] ($ (module7) + (-1.75cm,1ex) $) -- ($ (module7) + (-1.75cm-0.5cm-5ex-0.5cm-1.5cm,1ex) $);

\draw [<-, thick, KIT_gruen_1] ($ (module1) + (-1.75cm,-1ex) $) -- ($ (module1) + (-1.75cm-0.5cm,-1ex) $);
\draw [<-, thick, KIT_gruen_1] ($ (module3) + (-1.75cm,-1ex) $) -- ($ (module3) + (-1.75cm-0.5cm,-1ex) $);
\draw [<-, thick, KIT_gruen_1] ($ (module4) + (-1.75cm,-1ex) $) -- ($ (module4) + (-1.75cm-0.5cm,-1ex) $);
\draw [<-, thick, KIT_gruen_1] ($ (module5) + (-1.75cm,-1ex) $) -- ($ (module5) + (-1.75cm-0.5cm,-1ex) $);
\draw [<-, thick, KIT_gruen_1] ($ (module7) + (-1.75cm,-1ex) $) -- ($ (module7) + (-1.75cm-0.5cm,-1ex) $);

\node at ($ (module1) + (-1.75cm-0.5cm-5ex-0.5cm-0.5cm,1ex-1.75ex) $) {\textit{define}};
\node at ($ (module1) + (-1.75cm-0.5cm-5ex-0.5cm-0.5cm,1ex-1.75ex-2.25ex) $) {\textit{arguments}};

\draw [rounded corners, KIT_gruen_1, fill=white] ($ (module1) + (-1.75cm-0.5cm-5ex,2.5ex) $) rectangle node{\rotatebox{90}{Settings}} ($ (module8) + (-1.75cm-0.5cm,-2.5ex) $);

\draw [<-, thick, KIT_gruen_1] ($ (module4) + (-1.75cm-0.5cm-2.5ex-2.5ex,-2.5ex-0.15cm) $) -- ($ (module4) + (-1.75cm-0.5cm-5ex-2cm,-2.5ex-0.15cm) $);

\draw [rounded corners, KIT_gruen_1, fill=KIT_gruen_5] ($ (module1) + (-1.75cm-0.5cm-5ex-0.5cm-1.5cm-5ex-0.5cm-5ex,2.5ex) $) rectangle node{\rotatebox{90}{Program Arguments, JSON Configuration}} ($ (module8) + (-1.75cm-0.5cm-5ex-0.5cm-1.5cm-5ex-0.5cm,-2.5ex) $);

\draw [->, thick, KIT_gruen_1] ($ (module4) + (-1.75cm-0.5cm-5ex-0.5cm-1.5cm-5ex-0.5cm,-2.5ex-0.15cm) $) -- ($ (module4) + (-1.75cm-0.5cm-5ex-0.5cm-1.5cm-5ex,-2.5ex-0.15cm) $);

\draw [rounded corners, KIT_rot_1, fill=white] ($ (module1) + (1.75cm+0.5cm+5ex,2.5ex) $) rectangle node{\rotatebox{90}{Plot Data}} ($ (module8) + (1.75cm+0.5cm,-2.5ex) $);

\draw [<->, thick, KIT_rot_1] (module1) -- ($ (module1) + (1.75cm+0.5cm,0cm) $);
\draw [<->, thick, KIT_rot_1] (module3) -- ($ (module3) + (1.75cm+0.5cm,0cm) $);
\draw [<->, thick, KIT_rot_1] (module4) -- ($ (module4) + (1.75cm+0.5cm,0cm) $);
\draw [<->, thick, KIT_rot_1] (module5) -- ($ (module5) + (1.75cm+0.5cm,0cm) $);
\draw [<->, thick, KIT_rot_1] (module7) -- ($ (module7) + (1.75cm+0.5cm,0cm) $);
\end{tikzpicture}
\caption{Structure of the \harryplotter framework.
The processing of input modules is followed by the run of analysis modules and concluded by the execution of plot modules.
The configuration via a command line interface or via JSON files is parsed by an argument parser whose arguments are defined by the individual modules.}
\label{figure_artus_harry_plotter}
\end{figure}

The structure of the framework is illustrated in Figure~\ref{figure_artus_harry_plotter}.
It follows the idea of processors in the \artus framework.
Three kinds of processor modules are executed subsequently:
\begin{enumerate}
\item One or, in special cases, more input modules are run to read data from an input source.
The input format is defined by the input module.
The most commonly used input module is the one that reads histograms or other objects from ROOT files.
\item An arbitrary number of analysis modules is appended.
They can perform calculations based on the inputs or the results of previously run analysis modules.
\item One or, in special cases, multiple plot modules are executed in the last step.
Their task is to output the processed data in the specified format.
In most cases, the output is a plot, but it is also possible to store the objects in new ROOT files.
\end{enumerate}
All processors share a common plot data object and possible meta-data where they can add new information and read the existing one.
For example, the input modules add histograms read from input files and the plot modules access these histograms.

The configuration is provided by an argument parser that can handle both arguments from a command line interface and content from JSON configuration files or Python dictionaries.
This makes it possible to either use the \harryplotter executable configured via the command line or to call the main function within another Python script.
The executable can perform one single plot per run.
For the execution within a script, functions exist to create multiple plots in parallel.

The parsing of the arguments is performed in two steps:
First, the sequence of modules to be executed is determined.
These modules are then initialized and can define additional arguments needed for their configuration, such that every module can manage its own settings but can also access the settings of the entire program.

\section{\texorpdfstring{A Real World Application -- the $\PH\to\Pgt\Pgt$ Analysis}{A Real World Application - the H to Tau Tau Analysis}}
\label{section_artus_example_htt}

Currently, the \artus framework is used for various user analyses of CMS data, ranging from jet energy calibration and QCD studies to Higgs boson measurements and searches for physics beyond the Standard Model.
As an example, the technical aspects of the $\PH\to\Pgt\Pgt$ analysis are explained below.

The $\PH\to\Pgt\Pgt$ analysis makes use of skimmed data sets in the \kapa format.
The \kapa framework~\cite{kappa} is designed for skimming CMS data into small n-tuples that can be analyzed independently of the CMSSW software.
The only dependence is the ROOT framework.
The n-tuples store high-level information such as four-momentum vectors, collections of leptons and vertices.
Similar to the \artus framework, \kapa aims to satisfy general analysis needs.

A large part of the \artus software is specialized on the analysis of \kapa n-tuples\footnote{\url{https://github.com/artus-analysis/Artus/tree/master/KappaAnalysis}}.
An event provider is available for reading \kapa n-tuples.
The event class contains members for all important physics objects used in analyses.
The branches of the n-tuples, that should be read in from the input files, are configurable.
This reduces the amount of information transferred from the input files and therefore accelerates the analysis significantly.

The structure of the pipelines is chosen such that a separate pipeline exists for each decay channel.
Each of the pipelines is duplicated and modified in order to implement the different settings for shifts of quantities affected by systematic uncertainties.
For example, the \Pgt energy scale is shifted up and down with respect to the nominal value.
The change of the \Pgt four-momentum is then propagated through the entire analysis and the effect on the final result is studied.
For this, each pipeline for decay channels involving hadronically decaying \Pgt leptons is duplicated three times and the setting for the \Pgt energy scale shift is modified accordingly.
The global pipeline is used to perform actions that are common for all channels.
This is for example the identification and correction of jets as well as the filtering of valid data events.

The producers and filters are implemented to perform small tasks of the analysis.
Main producers are the ones for the identification and the selection of valid leptons and jets.
Further producers perform the matching with trigger or generator objects and perform the splitting into decay channels or the categorization.
Also the mass reconstruction is done in a separate producer.
The selection of events according to high level trigger decisions as well as the selection of decay channels are examples for filters.
The filters are executed as early as possible in the chain of processors in order to avoid running subsequent processors without real need.

\begin{table}[!ht]
\centering
\begin{tabular}{lrr}
\toprule
\textbf{Pipelines}                                & \textbf{Runtime / \si{min}} & \textbf{Event Rate / \si{s^{-1}} } \\ \midrule
 0                                                & 5:12                             & 1470 \\ \midrule
 1 ($\Pe\Pgm$)                                    & 6:31                             & 1174 \\
 1 ($\Pe\Pgt$)                                    & 7:26                             & 1028 \\
 1 ($\Pgm\Pgt$)                                   & 7:14                             & 1052 \\
 1 ($\Pgt\Pgt$)                                   & 6:07                             & 1247 \\ \midrule
 2 ($\Pgm\Pgt$, $\Pe\Pgt$)                        & 8:32                             & 896 \\
 3 ($\Pgm\Pgt$, $\Pe\Pgt$, $\Pe\Pgm$)             & 9:54                             & 767 \\
 4 ($\Pgm\Pgt$, $\Pe\Pgt$, $\Pe\Pgm$, $\Pgt\Pgt$) & 10:01                            & 758 \\
\bottomrule
\end{tabular}
\caption[Measurement of the runtime of the baseline $\PH\to\Pgt\Pgt$ analysis.]{Measurement of the runtime of the baseline $\PH\to\Pgt\Pgt$ analysis.
The analysis is performed on 10 files with a total size of 6.1~GB containing 452229 simulated $\PZ+\text{jet}$ events.
The measurement is performed with different numbers of pipelines, where each pipeline implements the analysis of one sub-channel.
The values are compared with the execution without any pipelines and without global processors.
It is clearly visible, that the time needed for reading the inputs amounts to a large part of the overall runtime and that the simultaneous execution of multiple pipelines can reduce the runtime significantly compared to the case where different configuration are performed sequentially including separate IO operations.}
\label{table_artus_runtime_comparison}
\end{table}

Table~\ref{table_artus_runtime_comparison} shows a runtime measurement for the baseline part of the $\PH\to\Pgt\Pgt$ in the four main decay channels ($\Pgm\Pgt$, $\Pe\Pgt$, $\Pe\Pgm$ and $\Pgt\Pgt$).
It illustrates the main advantage of the \artus structure: sub-analyses with different sets of configurations can be performed in the same run avoiding the multiple I/O overhead when the configurations would be executed sequentially in different runs.

The outputs are flat n-tuples suitable for performing the background modeling step which is done in \harryplotter.
Histograms are read while the final event selection is applied.
As an example, the selection on the transverse mass is applied on this level because events from control regions need to be preserved for the background modeling step.
For each background process, an analysis module has been developed performing the estimation of the samples.
Most of them perform calculations on multiple histograms, defining the yield and the shape of the final quantity in signal and control regions.
They add a single histogram to the plot data object containing the final estimation for a given sample.
Similarly, ratios comparing the compatibility of data and the simulation for the sub-plots are determined and added this way.

The post-processing step either outputs plots or writes the histograms to ROOT files and passes them over to the limit calculation tools.

\section{Conclusion}
\label{section_conclusion}

The \artus analysis framework presented in this paper offers a comprehensive and well-integrated solution for event-based data analyses in the HEP domain.
In conjunction with the post-processing toolkit \harryplotter, all typical data analysis needs are covered.

The modular and structured manner, in which the processing required for an analysis, can be expressed within the \artus framework, allowed the involved research groups to share and re-use source code on a scale not possible before.
The concept underlying the \artus framework also enabled sharing program code between analyses and even across different universities and research centers.

Especially new students joining the analysis groups can contribute very quickly.
Having understood general concept and structure of an \artus analysis, they can contribute by writing their own producers or filters while the full complexity of the event-processing is taken care of by the \artus framework.
Using the \artus framework, it is much less work to set up a new analysis because a rich library of processors, filters and consumers already exists and can be reused for new analyses.

\end{document}